\documentclass{aa}
\usepackage{graphicx}
\usepackage{txfonts}
\usepackage{hyperref}
\usepackage{orcidlink}
\usepackage[normalem]{ulem}
\usepackage{pifont}

\hypersetup{colorlinks = true, citecolor = {blue}, urlcolor= {blue}}

\def\gapprox{\;\rlap{\lower 3.0pt                       
        \hbox{$\sim$}}\raise 2.5pt\hbox{$>$}\;}
\def\lapprox{\;\rlap{\lower 3.1pt                       
        \hbox{$\sim$}}\raise 2.7pt\hbox{$<$}\;}

\newcommand{\be}{ \begin{equation} }
\newcommand{\ee}{\end{equation}}

\newcommand{\ben}{\begin{enumerate}}
\newcommand{\een}{\end{enumerate}}

\renewenvironment{thebibliography}[1]{\begin{oldthebibliography}{#1}\setlength{\parskip}{0ex}\setlength{\itemsep}{0ex}}{\end{oldthebibliography}}

\usepackage{diagbox}
\makeatletter
\renewcommand*\aa@pageof{, page \thepage{} of \pageref*{LastPage}}
\makeatother

\usepackage[dvipsnames]{xcolor}

\usepackage{booktabs}

\titlerunning{The effect of triaxial galaxy shapes on the dynamics of triple supermassive black holes in a cosmological context}
\authorrunning{N. Saha et al.}

\begin{document}

    \title{The effect of triaxial galaxy shapes on the dynamics of triple supermassive black holes in a cosmological context}

   \author{
	Navonil Saha\inst{1}\thanks{Fellow of the International Max Planck Research School for Astronomy and Cosmic Physics at the University of Heidelberg (IMPRS-HD)}
    \and Margarita~Sobolenko\inst{2,3}
    \and
    Peter Berczik\inst{2,3}
    \and
    Andreas Just\inst{1}
    \and
    Fazeel~Mahmood~Khan \inst{4,5}
    }

\institute{
{Zentrum f\"ur Astronomie der Universit\"at Heidelberg, Astronomisches Rechen-Institut, M\"{o}nchhofstr. 12-14, 69120, Heidelberg, Germany, \email{\href{mailto:navonil.saha@uni-heidelberg.de}{navonil.saha@uni-heidelberg.de}}}
\and
{Main Astronomical Observatory, National Academy of Sciences of Ukraine, 27 Akademika Zabolotnoho St, 03143 Kyiv, Ukraine}  
\and
{Nicolaus Copernicus Astronomical Centre, Polish Academy of Sciences, ul. Bartycka 18, 00-716 Warsaw, Poland}
\and
{New York University Abu Dhabi, PO Box 129188, Abu Dhabi, United Arab Emirates}
\and
{Center for Astrophysics and Space Science (CASS), New York University Abu Dhabi
}
}
\date{Received 13 March 2026 / Accepted 22 June 2026}

  \abstract
   {

The hierarchical nature of galaxy formation in the Lambda cold dark matter ($\Lambda$CDM) cosmological framework model often leads to the presence of multiple supermassive black holes (SMBHs) in the galactic nuclei. The timescale over which galaxies merge plays a crucial role in shaping the dynamical evolution and the merger dynamics of their central SMBHs. While binary SMBH evolution has been extensively studied, the long-term dynamics of triple SMBH systems, especially in realistic, nonspherical galactic potentials, still remain less understood. In this work, we investigated the role of triaxiality in shaping the dynamical evolution of three SMBH triple systems taken from the ROMULUS25 cosmological simulation embedded in triaxial stellar backgrounds to find common dynamical evolution patterns and estimate typical coalescence times using high-resolution gravitodynamical $\textit{N}$-body simulations. We explored a range of orbital configurations and host galaxy shapes with initial conditions from the ROMULUS25 data and tracked the orbital evolution from the galactic inspiral to the formation of hard binaries at sub-parsec separations and used the observed hardening rates to estimate the time of coalescence. In all cases, the two heaviest black holes form an efficiently hardening binary, which merges within the Hubble time, while the third black hole (BH) either forms a stable hierarchical triple system with the heavier binary or remains on a wide galactic orbit. Finally, we analyzed the triaxiality of the galactic remnant from our simulations and conclude that the initial triaxial shape of the galaxies does not significantly change the final dynamical outcome of the triple systems.

   }


   \maketitle
%

\section{Introduction}
During their lifetimes, most massive galaxies undergo various major merger events. This has been the basis of the hierarchical cosmic structure formation in the Lambda cold dark matter ($\Lambda$CDM) cosmology \citep{1970ApJ...162..815P, 1978MNRAS.183..341W, 1991ApJ...379...52W}, which has already been verified through several observational and numerical studies \citep[see, e.g.,][]{2006ApJ...652..270B, 2006ApJ...636L..81N, 2010ApJ...709.1018V, 2021MNRAS.501.3215O}. It is also well established that most of the elliptical and spiral galaxies contain a central supermassive black hole \citep[SMBH;][]{Kormendy+13,gra16}. The empirical scaling relations suggest that the galaxies and their central SMBHs grow together. For example, observational studies have revealed tight empirical correlations between the masses of SMBHs and the properties of their host galaxies, most notably the $M_\mathrm{BH}$–$\sigma$ relation, which links the SMBH mass ($M_\mathrm{BH}$) to the stellar velocity dispersion ($\sigma$) of the galactic bulge \citep{2000ApJ...539L...9F, 2000ApJ...539L..13G}. 
Similarly, the $M_\mathrm{BH}$–$M_{\rm bulge}$ and $M_\mathrm{BH}$–$L_{\rm bulge}$ relations connect SMBH mass to the bulge stellar mass and luminosity, respectively \citep{1998AJ....115.2285M, 2004ApJ...604L..89H}.
These scaling relations point to a strong coupling between SMBH growth and the assembly of their host galaxies. The tightness and apparent universality of correlations between SMBH mass and bulge properties suggest that the processes regulating star formation and bulge buildup are linked to those governing SMBH mass growth via accretion and mergers, consistent with a coevolutionary picture.

The dynamical evolution of the SMBH binaries following a galaxy merger is commonly divided into three distinct stages governed by different physical mechanisms \citep{1980Natur.287..307B}. First, in the post-merger phase of galaxy mergers, the SMBHs sink to the center of the galaxy through dynamical friction against the background of stellar particles \citep{1943ApJ....97..255C}. The gravitational drag decelerates the SMBHs, causing them to lose orbital energy and angular momentum. This process operates efficiently from kilo-parsec scales down to separations on the order of a few parsecs, where the two SMBHs become gravitationally bound. Once a bound binary forms, further orbital decay is dominated by interactions with the surrounding stellar environment. This is known as the stellar hardening phase. During this phase, stars passing close to the binary extract energy and angular momentum through three-body interactions (the gravitational slingshot mechanism). This causes the semimajor axis to shrink and thus form a hard binary \citep{1996NewA....1...35Q, 2010ApJ...719..851S}. 
The efficiency of this mechanism depends on the rate at which stars are supplied to the loss cone \citep{Baile15} in phase space and without an efficient replenishment mechanism of the loss-cone orbits, the dynamical evolution of the SMBH binary can stall at parsec scales, which gives rise to the infamous final parsec problem \citep{2003ApJ...596..860M, 2005ApJ...633..680B}.
Various analytical and numerical studies suggest that stellar dynamics alone can accommodate the SMBH binary evolution in the hardening phase, ensuring an efficient transition to the gravitational wave (GW) dominated regime \citep{1983AJ.....88.1269H, 1991Natur.354..212E, 2006ApJ...651..392S, 2008ApJ...686..432S, 2011MNRAS.411..653J, 2012ApJ...749..147K, 2013ApJ...773..100K, 2016ApJ...828...73K, 2020MNRAS.493L.114B, 2022MNRAS.511.4753G}.
Finally, when the separation of the black holes (BHs) falls below sub-parsec scales, GW emission dominates the energy loss, rapidly driving the binary to coalescence and merger \citep{1963PhRv..131..435P, 1964PhRv..136.1224P}. Forthcoming observations with the ESA--NASA Laser Interferometer Space Antenna (LISA) will open a new window into the merger history of SMBHs, enabling detections across much of cosmic time, including events at high redshifts \citep{ama23,colpi24}.

The timescale from the merger of galaxies to the final coalescence of their central SMBHs can vary significantly, depending on the remnant’s matter distribution and the binary’s orbital parameters, ranging from tens of megayears (Myr) to a few gigayears (Gyr) or longer \citep{2012ApJ...749..147K,Khan+15, holley+15, 2014SSRv..183..189C, 2022MNRAS.511.4753G,khan+24, 2025A&A...703A..86F}. During this evolution, it is possible that a third galaxy with its SMBH merges with the merger remnant, which already contains a SMBH binary and creates a triple SMBH system. This has been confirmed by various observations of triple active galactic nuclei (AGN) systems over the last few years \citep{2014Natur.511...57D, 2017ApJ...851L..15K, 2019ApJ...883L...7L,2019ApJ...887...90L, 2019A&A...630L...5G, 2019ApJ...883..167P,  2020A&A...633A..79K, 2021A&A...651L...9Y, 2022ApJ...934...89P, 2025PhRvD.111j3016L}.

Hence, it becomes very crucial to study how such a triple SMBH system evolves throughout the course of the galactic merger and how the presence of a third black hole can modify the dynamical evolution as compared to a binary SMBH merger. On one hand, the SMBHs either form subsequent binaries orbited by a distant tertiary black hole in a stable hierarchical configuration, which eventually merge on conventional timescales \citep{2007MNRAS.377..957H, 2018MNRAS.473.3410R}, or they undergo a chaotic evolution, where one or more black holes are ejected from the galactic nucleus \citep{2006ApJ...651.1059I, 2012MNRAS.422.1306K}. We already know that cosmological simulations are not capable of resolving the dynamics of the SMBH down to parsec scales, and as a result the long-term dynamics of SMBHs triples remain poorly understood. Hence, high-resolution follow-up simulations are required to correctly model the dynamics of the SMBHs in the galactic nuclei. In the context of SMBH triples, studies have been conducted using both few-body approaches \citep{1989CeMDA..46..277V, 1996MNRAS.278..186V, 2002ApJ...578..775B,2007MNRAS.377..957H,2016MNRAS.461.4419B, 2018MNRAS.477.3910B,2018MNRAS.473.3410R} and $N$-body simulations \citep{2006ApJ...651.1059I, 2008arXiv0801.0859I, 2010MNRAS.402.2308A, 2012MNRAS.422.1306K,2021A&A...649A..41A, 2021ApJ...912L..20M, 2022ApJ...929..167M, 2023A&A...678A..11K, 2024A&A...687L..18B, 2025A&A...703A..86F}. Among these studies, only \cite{2021A&A...649A..41A}, \cite{2021ApJ...912L..20M, 2022ApJ...929..167M}, \cite{2023A&A...678A..11K}, and \cite{2025A&A...703A..86F} used cosmological simulations as the source for their initial conditions.

In \cite{2023A&A...678A..11K}, initial conditions of triple SMBH systems and their host galaxies were extracted from the galaxy merger trees. The authors find that the triple interactions play a crucial role in the outcome of the dynamical evolution of the SMBHs. In these simulations, spherical galaxy systems (following the same procedure as \citealt{2021A&A...649A..41A}) were used with the initial conditions taken from the ROMULUS25 cosmological simulation \citep{2017MNRAS.470.1121T}, but neglecting a more realistic triaxial galaxy structure. It turned out that in all models, the two most massive SMBHs form an efficiently hardening binary and merge first within fractions of the Hubble time. On the other hand, the lightest one either gets ejected, forms a stable hierarchical triple system with the heavier binary, or remains on a wide galactic orbit. 

In this present work, we extended these models 
of triple SMBHs by investigating the role of the triaxial shape of host galaxies in shaping the dynamical evolution of three SMBH systems. Our goal was to find the common orbital evolution pattern of the triple system and predict the typical coalescence timescale using high-resolution $N$-body simulations. In Sect.~\ref{Triaxial Galaxies}, we describe the galaxies' triaxial shape. In Sect.~\ref{Model Setup}, we detail the modeling of our triple system and the initial conditions. In Section~\ref{Results}, we discuss our results of the triaxial triple merger runs. Lastly, in Sect.~\ref{Discussion and conclusions} we conclude our discussions. Throughout this paper we assume a $\Lambda$CDM cosmology with a Hubble constant of $H_{0}=67.74$~km~s$^{-1}$~Mpc$^{-1}$, $\Omega_{\rm M}=0.30$, and $\Omega_{\Lambda}=0.69$ \citep{2016A&A...594A..13P}

\section{Triaxial galaxies} \label{Triaxial Galaxies}

The intrinsic shape of galaxies plays a crucial role in the dynamical evolution of SMBHs. From various cosmological simulations, it is already known that most of the host galaxies are already triaxial in shape \citep{1991ApJ...383..112F, 2002ApJ...574..538J, 2006MNRAS.367.1781A, 2011MNRAS.416.1377V}. Even merger remnants of initially spherical galaxies, which begin with isotropic stellar distributions and central potentials symmetric along all axes, develop weak triaxiality as a result of dynamical instabilities and violent relaxation during the merging process \citep{1992ApJ...400..460H, 1992ApJ...393..484B, 2003ApJ...597..893N,2011ApJ...732...89K,Gualandris+12, 2018MNRAS.477.2310B}. Triaxial galaxies deviate from spherical symmetry and are described by three unequal principal axes $(\mathrm{a}\ge \mathrm{b}\ge \mathrm{c})$, corresponding to the major, intermediate, and minor axes, respectively, rather than a single symmetry axis. A quantitative measure of the intrinsic shape of a stellar system is given by the triaxiality parameter ($\mathrm{T}$)  as follows \citep{1978MNRAS.183..501B, 1991ApJ...383..112F}:
\begin{equation}
    \mathrm{T}=\frac{\mathrm{a}^2-\mathrm{b}^2}{\mathrm{a}^2-\mathrm{c}^2} = \frac{1-(\mathrm{b}/\mathrm{a})^2}{1-(\mathrm{c}/\mathrm{a})^2},\ \ \text{where }0\leq \mathrm{T} \leq1.
\end{equation}
Here, $\mathrm{T}=0$ corresponds to a perfectly oblate spheroid where $\mathrm{a}=\mathrm{b}>\mathrm{c}$, and $\mathrm{T}=1$ to a perfectly prolate spheroid, where $\mathrm{a}>\mathrm{b}=\mathrm{c}$. All other intermediate values, $0<\mathrm{T}<1$, represent truly triaxial structures. The triaxiality parameter is also directly related to the axis ratios $\mathrm{b/a}$ and $\mathrm{c/a}$, where $\mathrm{b/a}$ quantifies how much the intermediate axis differs from the major axis (i.e., how elongated the galaxy is in the $x-y$ plane) and $\mathrm{c/a}$ quantifies how much the minor axis differs from the major axis (i.e., how flattened the galaxy is along the $z$-axis).

In the dynamical evolution of the SMBHs during the merger process, the orbits of the triaxial galaxy shape also differ from those of the spherical galaxies. Triaxial potentials usually give rise to centrophilic orbits, which can efficiently drive stars into the galactic center helping replenish the loss cone and hence solve the final parsec problem \citep{2006ApJ...642L..21B, 2011ApJ...732...89K, 2011ApJ...732L..26P, 2013ApJ...773..100K, 2016ApJ...828...73K, 2017MNRAS.464.2301G}. In this current work, we investigate the effect of the triaxiality of progenitor galaxies on the formation and evolution of the triple SMBH systems. 

\section{Model setup} \label{Model Setup}

The ROMULUS25 cosmological simulations focus on SMBH physics, where SMBHs are freely moving and subjected to sub-grid dynamical friction \citep{2015MNRAS.451.1868T}. It also includes improved models for SMBH seeding, accretion, and feedback. For a more detailed explanation, refer to \cite{2017MNRAS.470.1121T}. The main run of the simulation is ROMULUS25, which spans a 25~cMpc cube run to $z=0$ with a Plummer softening of  0.25~kpc and a particle mass resolution of $m_\mathrm{DM}=3.39\times10^5M_\odot$. The SMBHs merge when their separation $\Delta R_\mathrm{BH}$ falls below 0.7~kpc; hence, the proper SMBH orbital evolution cannot be traced down to the actual merger.

To properly resolve the SMBH dynamics down to sub-parsec scale, we used the same triple systems identified by \cite{2023A&A...678A..11K}. They used specific selection criteria where, the remnant of a dual merger had to undergo a subsequent merger with a third black hole within the following 1~Gyr, and selected three triple SMBH systems labeled A, B, and C. In system A, three galaxies hosting $10^7-$$10^8M_\odot$ SMBHs merge simultaneously. System B involves a merger between two galaxies, one containing a pair of $10^7M_\odot$ SMBHs from a previous merger and the other hosting a $5\times10^7M_\odot$ SMBH. Finally, in system C, two smaller $10^7M_\odot$ SMBHs orbit a central $10^9M_\odot$ SMBH. Hence, our initial conditions represent three, two, and one galaxies for systems A, B, and C, respectively.

\subsection{Triaxial initial conditions} \label{Triaxial initial conditions}
To create different models of triaxial initial conditions with dark matter, gas and stars, we used the spherical galaxy models from \cite{2023A&A...678A..11K}. We then fitted these galactic components with a Dehnen profile \citep{1993MNRAS.265..250D}: 
\begin{align}
    \rho(r)&=\frac{(3-\gamma)M}{4\pi}\frac{r_{\rm s}}{r^\gamma\left(r+r_{\rm s}\right)^{4-\gamma}},\\
    m(r)&=M\left(\frac{r}{r+r_{\rm s}}\right)^{3-\gamma}.
\end{align}
Here, the equations give the density $\rho(r)$ and cumulative mass profiles $m(r)$ respectively, where $M$ is the total mass and $r_{\rm s}$ is the scale radius. The parameter $\gamma$ controls the inner slope of the stellar density profile, where a larger $\gamma$ corresponds to a steeper central cusp while a smaller $\gamma$ gives a shallower core-like center. To examine how changes in the central profile influence the results, model A was further divided into three stellar density profile variants labeled A.1, A.2, and A.3. In A.1, the stellar density has a flat low-density core ($\gamma = 0$). A.3 has $\gamma = 1$ with a cuspy, dense core. A.2 is intermediate, with $\gamma = 1$ but central density $\sim 3$ times lower than A.3 due to a larger scale radius. The three galaxies in A.2 have slightly higher stellar masses than the other models. 
The radial galactic Dehnen profile parameters were taken from \cite{2023A&A...678A..11K} for all five models and are given in Table~\ref{tab:IC_parameters}. The stellar density profiles were fitted outside $\sim$1 kpc and extrapolated inward analytically. Since the ROMULUS25 profiles become numerically unresolved inside a few gravitational softening lengths ($\epsilon = 0.25$ kpc), the inner logarithmic slope ($\gamma$) is not uniquely constrained by the cosmological simulation. We also rechecked the density and mass profiles for all models and obtain consistent results. The observed differences in the initial galactic density profiles are attributable to the varying stellar components of the respective models. 

We added the triaxial shape taken from \cite{2016MNRAS.462..663B} to these galaxies using AGAMA galaxy modeling software \citep{2019MNRAS.482.1525V} to generate the three-component $N$-body models in dynamical equilibrium. Here, the potential of the SMBHs was not taken into account to allow for the formation of a steeper central cusp within the influence radius of the SMBHs. Hence, we added the black holes after the galaxies had been generated and allowed the systems to evolve for a few dynamical timescales of the nucleus in order to create an additional stellar cusp around the central potential of the SMBHs \citep{1995ApJ...440..554Q}. The initial positions and velocities of the SMBHs were taken from the ROMULUS25 orbital data. The initial conditions for systems A and B were created with $\approx 25\times 10^6$ particles while for system C we had $\approx 19\times 10^6$ particles for sufficient interaction of the particles during the early evolution phase of the triples. The axis ratios $(\mathrm{b}/\mathrm{a}\ \text{and}\ \mathrm{c}/\mathrm{a})$ were chosen such that they showed a triaxial structure for all the components with similar triaxial values across all the models. 

\begin{table}
    \caption{Galactic parameters for the triaxial initial conditions.}
    {\small
    \begin{tabular}{p{0.05 \textwidth} p{0.05 \textwidth}p{0.05 \textwidth} p{0.03 \textwidth}p{0.03 \textwidth}p{0.03 \textwidth}p{0.03 \textwidth}p{0.03 \textwidth}}
    \toprule
    Galaxy & Species & Property & A.1 & A.2 & A.3 & B & C\\
    \midrule
        1 & DM & $r_{\rm s}$ & 33.33 & 33.3 & 33.3 & 39.8 & 74.6 \\
         & & $M$ & 216.9 & 216.9 & 216.9 & 109.0 & 892.9 \\
         & & $\mathrm{b}/\mathrm{a}$ & 0.9 & 0.9 & 0.9 & 0.9 & 0.9 \\
         & & $\mathrm{c}/\mathrm{a}$ & 0.7 & 0.7 & 0.7 & 0.7 & 0.7 \\
         \rule{0pt}{3ex}
         & Stars & $r_{\rm s}$ & 0.7 & 2.2 & 1.3 & 0.4 & 4.4\\
         & & $\gamma$ & 0 & 1 & 1 & 1.4 & 1.9\\
         & & $M$ & 6.5 & 10.4 & 7.1 & 5.2 & 2.6\\
         & & $\mathrm{b}/\mathrm{a}$ & 0.8 & 0.8 & 0.8 & 0.8 & 0.8 \\
         & & $\mathrm{c}/\mathrm{a}$ & 0.7 & 0.7 & 0.7 & 0.7 & 0.7 \\
         \rule{0pt}{3ex}
         & Gas & $r_{\rm s}$ & 49.1 & 49.1 & 49.1 & 40.0 & 149.7 \\
         & & $M$ & 34.3 & 34.3 & 34.3 & 6.2 & 198.7 \\
         & & $\mathrm{b}/\mathrm{a}$ & 0.9 & 0.9 & 0.9 & 0.9 & 0.9 \\
         & & $\mathrm{c}/\mathrm{a}$ & 0.8 & 0.8 & 0.8 & 0.8 & 0.8 \\
    \midrule
        2 & DM & $r_{\rm s}$ & 18.8 & 18.8 & 18.8 & 5.5 & - \\
         & & $M$ & 64.9 & 64.9 & 64.9 & 10.4 & -\\
         & & $\mathrm{b}/\mathrm{a}$ & 0.8 & 0.8 & 0.8 & 0.8 & - \\
         & & $\mathrm{c}/\mathrm{a}$ & 0.6 & 0.6 & 0.6 & 0.6 & - \\
         \rule{0pt}{3ex}
         & Stars & $r_{\rm s}$ & 1.2 & 3.8 & 2.2 & 2.31 & -\\
         & & $\gamma$ & 0 & 1 & 1 & 2 & -\\
         & & $M$ & 4.5 & 9.1 & 5.2 & 2.3 & -\\
         & & $\mathrm{b}/\mathrm{a}$ & 0.8 & 0.8 & 0.8 & 0.8 & - \\
         & & $\mathrm{c}/\mathrm{a}$ & 0.7 & 0.7 & 0.7 & 0.7 & - \\
         \rule{0pt}{3ex}
         & Gas & $r_{\rm s}$ & 9.1 & 9.1 & 9.1 & 3.2 & -\\
         & & $M$ & 2.3 & 2.3 & 2.3 & 0.7 & -\\
         & & $\mathrm{b}/\mathrm{a}$ & 0.9 & 0.9 & 0.9 & 0.9 & - \\
         & & $\mathrm{c}/\mathrm{a}$ & 0.8 & 0.8 & 0.8 & 0.8 & - \\
    \midrule
        3 & DM & $r_{\rm s}$ & 3.1 & 3.1 & 3.1 & - & -\\
         & & $M$ & 10.4 & 10.4 & 10.4 & - & - \\
         & & $\mathrm{b}/\mathrm{a}$ & 0.8 & 0.8 & 0.8 & - & - \\
         & & $\mathrm{c}/\mathrm{a}$ & 0.6 & 0.6 & 0.6 & - & - \\
         \rule{0pt}{3ex}
         & Stars & $r_{\rm s}$ & 0.4 & 1.0 & 0.6 & - & -\\
         & & $\gamma$ & 0 & 1 & 1 & - & - \\
         & & $M$ & 3.0 & 3.5 & 3.1 & - & -\\
         & & $\mathrm{b}/\mathrm{a}$ & 0.8 & 0.8 & 0.8 & - & - \\
         & & $\mathrm{c}/\mathrm{a}$ & 0.7 & 0.7 & 0.7 & - & - \\
         \rule{0pt}{3ex}
         & Gas & $r_{\rm s}$ & - & - & - & - & -\\
         & & $M$ & - & - & - & - & -\\
    \bottomrule
    \end{tabular}}
    \tablefoot{Triaxial Dehnen parameters for the different matter components in the galaxies of our models. The scale radii $r_{\rm s}$ are given in kiloparsec, the mass values $M$ in $10^{10} M_\odot$. The triaxial axis-ratios are given as $\mathrm{b}/\mathrm{a}$ and $\mathrm{c}/\mathrm{a}$.}
    \label{tab:IC_parameters}
\end{table}
\label{subsec:Construction_of_IC}

For the initial conditions, we selected snapshots at 6.25~Gyr, 7.7~Gyr, and 9.49~Gyr cosmic time corresponding to redshifts of $z$\ =\ 0.91, 0.63, and 0.39 from ROMULUS25 for models A, B, and C, respectively. These correspond to the initial time $t_\mathrm{start}$ for our simulation, as described in \cite{2023A&A...678A..11K}. We ended our simulation at the time corresponding to $t_\mathrm{end}$ for our $N$-body runs. The masses of the black holes in ROMULUS25 also undergo accretion, since the $N$-body codes do not account for the SMBH growth; therefore, to account for this effect, the SMBH mass for all the triple models was set to their values just immediately before their last merger, as given in Table~\ref{tab:mass_of_SMBH}. The most massive black hole is labeled as BH1, followed by BH2 for the intermediate one, and finally BH3 for the lightest SMBH. 

\subsection{Simulation techniques}
For the initial interaction phase, we began our simulation with the fast oct-tree code \texttt{Bonsai2}, a custom version of the original \texttt{Bonsai}\footnote{https://github.com/treecode/Bonsai} code \citep{2012JCoPh.231.2825B}. It was used to model the early evolution of the global galaxy merger and the subsequent decay of the SMBHs toward the center of the merger remnant through dynamical friction. The symplectic integrator in \texttt{Bonsai2} accurately follows the orbital evolution during the galactic inspiral phase provided that the SMBH pairs remain unbound at large separations in the order of 0.1-100 kpc. Once the binary begins to harden and the separation of the inner black hole pair falls below $\sim$100 pc, the uniform timestep increases the risk of improperly resolving close pericenter passages, resulting in a time resolution problem.

\begin{figure*}
    \centering
    \includegraphics[width=\textwidth]{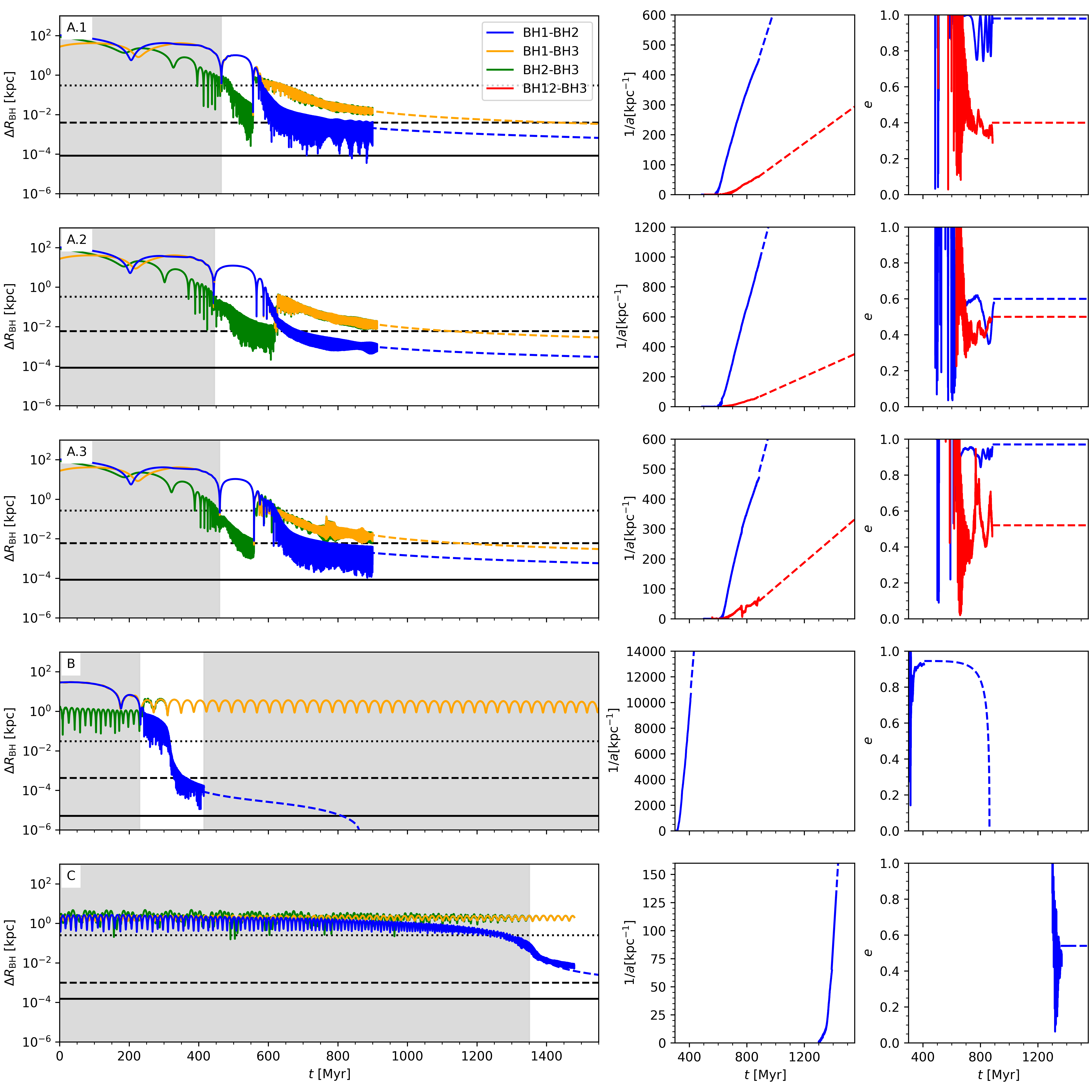}
    \caption{Black hole orbital parameters across the different simulations. The color coding refers to different SMBH pairs. The blue lines indicate the orbital parameters around the inner BH1-BH2 binary (the two most massive BHs). The red lines indicate the orbital parameters for BH3 with respect to the inner binary. The background colors mark the codes used during the particular simulation period (light gray for \texttt{Bonsai2}, white for \texttt{$\varphi$-GPU}). Left: Separation of different SMBH pairs. Middle: Evolution of the inverse semimajor axis. Right: Eccentricity evolution for bound SMBH systems. The dotted, dashed, and solid black lines represent the influence radius, $a_{h}$, and 1000 times the Schwarzschild radius of BH1, respectively.}

    \label{fig:triple_evolution}
\end{figure*}

At this point, we shifted to the \texttt{$\varphi$-GPU} \footnote{https://github.com/berczik/phi-GPU-mole} direct $N$-body code \citep{2011hpc..conf....8B, 2022A&A...665A..86B, 2021A&A...652A.134S, 2022MNRAS.517.1791S} with a fourth-order Hermite integrator and an individual block timestep scheme. Both codes were accelerated by performing the computations on graphics processing units (GPUs). For all models, we used \texttt{$\varphi$-GPU} during the late hardening phase of the SMBHs to accurately resolve the three-body dynamics close to the black holes until a bound, hardening binary was formed. The sequence of code usage is shown in Fig.~\ref{fig:triple_evolution}. Since \texttt{$\varphi$-GPU} can currently handle up to $\sim10$~million particles, we performed a radial cut and selected the particles within the inner 10~kpc around the center of the most massive galaxy containing all the SMBHs, reducing the total number of particles to 4.8 million for system A, 7.2 million for system B, and 7.9 million for system C. We adopted the same global softening of $\epsilon=10^{-3}$~kpc in \texttt{Bonsai2}. After shifting to \texttt{$\varphi$-GPU}, we reduced it further to $\epsilon=10^{-4}$~kpc for general particle-particle interactions and $\epsilon=10^{-6}$~kpc for SMBH interactions.

For the late phase evolution with the direct code, we included post-Newtonian (PN) approximation \citep{2017KPCB...33...21S, 2021A&A...652A.134S, 2025A&A...703A..23B} up to order 2.5 in our models when the two SMBHs reached a separation of $\sim$ 1000$R_\mathrm{s}$ (where $R_\mathrm{s}$ is the Schwarzschild radius) and became a bound binary. We continued our runs until a stable SMBH configuration was reached, i.e., either a hard binary or a stable hierarchical triple system was formed with a constant hardening rate and no further interesting evolution was reported. 

In simulation B, we had to artificially merge the particles representing the BH1-BH2  pair, when they reached a separation of $\sim10^{-4}$ kpc in order to avoid the high numerical cost related to the direct integration using \texttt{$\varphi$-GPU}. We then used Eqs.~(\ref{1/a_petermathews}) and (\ref{e_petermathews}) (see Sect.~\ref{Dynamics of massive binaries}), to predict the orbits until their physical merger, while the evolution of the BH3 proceeded as usual using the tree code. In simulation C, all the SMBHs remained in the same halo for an extended period and showed no significant evolution due to the lower density of the system during the first 1~Gyr. Therefore, due to computational limitations, we stopped the simulation shortly after the inner pair began to form a bound binary with a constant stellar hardening.

\begin{table}
    \caption{Timescales and black hole masses in the different models used in our simulation.}
     {\small
    \begin{tabular}{p{0.1 \textwidth} p{0.045 \textwidth} p{0.045 \textwidth} p{0.045 \textwidth} p{0.045 \textwidth} p{0.045 \textwidth}}
    \toprule
    Property & A.1 & A.2 & A.3 & B & C \\
    \midrule
    $t_{\mathrm{start}}$, Gyr  & 6.25 & 6.25 & 6.25 & 7.77 & 9.49\\
    $t_{\mathrm{end}}$, Gyr & 7.15 & 7.16 & 7.15 & 9.27 & 10.97\\
    \midrule
    $m_{\mathrm{BH1}}$, $10^{7}\ M_\odot$ & 88.4 & 88.4 & 88.4 & 5.4 & 160.4 \\
    & (56.6) & (56.6) & (56.6) & (4.8) & (159.2)\\
    $m_{\mathrm{BH2}}$, $10^{7}\ M_\odot$ & 13.3 & 13.3 & 13.3 & 1.9 & 3.3\\
                     & (8.1) & (8.1) & (8.1) & (1.6) & (3.2)\\
    $m_{\mathrm{BH3}}$, $10^{7}\ M_\odot$ & 3.6 & 3.6 & 3.6 & 1.1 & 1.1\\
                     & (3.5) & (3.5) & (3.5) & (1.1) & (1.0)\\
    \bottomrule
    \end{tabular}}
    \tablefoot{The mass values in brackets correspond to the ROMULUS25 black hole masses at the initial time $t_{\mathrm{start}}$, while the values actually adopted in our simulation account for the accretion that occurs in ROMULUS25 over the triple's lifetime.}
    \label{tab:mass_of_SMBH} 
\end{table}

\subsection{Dynamics of massive binaries} \label{Dynamics of massive binaries}
The two SMBHs are considered to be bound if the separation between them falls below the influence radius $(r_\mathrm{infl})$. It is defined by the radius enclosing a stellar mass equal to twice the total mass of the SMBH binary \citep{2004cbhg.symp..263M}:  
\begin{equation}
      M_\star (r_{\mathrm{infl}})=2(m_\mathrm{BH1}+m_\mathrm{BH2}).
\end{equation} 
A SMBH pair is called “hard” when its binding energy per unit mass exceeds $\sim\sigma^2$ (where $\sigma$ is the velocity dispersion of the surrounding matter). Therefore, the binary is considered hard if its separation falls below
\begin{equation}
    a_h=\frac{Gm_\mathrm{BH2}}{4\sigma^2}.
\end{equation}
Once dynamical friction becomes ineffective at sub-parsec scales, three-body scattering between the binary and the stars dominates. The SMBHs then lose energy to these lighter stars and begin to harden at a roughly constant rate. This rate is obtained by fitting a straight line to the inverse semimajor axis and finding its slope, defined as $s_\mathrm{SH}=\frac{\mathrm{d}}{\mathrm{d}t}\left(\frac{1}{a}\right)$. The hardening rate is usually expressed in terms of a dimensionless hardening parameter $H$ \citep{1996NewA....1...35Q, 2001ApJ...556..245M}:
\begin{equation}
    H=\frac{\sigma}{G\rho}\frac{\mathrm{d}}{\mathrm{d}t}\left(\frac{1}{a}\right),
\end{equation}
where $\rho$ is the density of the stellar particles and $a$ is the Keplerian semimajor axis. \citet{2015MNRAS.454L..66S} concluded that $\rho$ and  $\sigma$ should be adopted at the influence radius of the SMBH binary. 

After our simulation reached $t_\mathrm{end}$, we took the orbitally averaged hardening rate due to GW emission assuming a constant eccentricity value at $t_\mathrm{end}$. We then added the stellar hardening estimated from our simulation and calculated the projected merger time of the hard binary by computing the time evolution of the semimajor axis 
$a$ and eccentricity $e$ using the empirical formulas by \citep{1963PhRv..131..435P, 1964PhRv..136.1224P}:
\begin{equation} \label{1/a_petermathews}
    \frac{\mathrm{d}}{\mathrm{d}t}\left(\frac{1}{a}\right)=s_\mathrm{SH}+\frac{64}{5}\frac{G^3\mu M^2}{c^5a^5} \frac{1+(73/24)e^2+(37/96)e^4}{(1-e^2)^{7/2}},
\end{equation}\vspace{-0.3cm}
\begin{equation} \label{e_petermathews}
    \frac{\mathrm{d}e}{\mathrm{d}t}=-\frac{304}{15}\frac{G^3\mu M^2}{c^5a^4}\frac{e+(121/304)e^3}{(1-e^2)^{5/2}}.
\end{equation}
Here, $\mu$ is the reduced mass of the binary and $M$ is its total mass. The second term in Eq.~(\ref{1/a_petermathews}) gives the energy loss by GW emission, and Eq.~(\ref{e_petermathews}) describes how the binary circularizes through GW emission.

\section{Results} \label{Results}
\subsection{Dynamical evolution of the triple systems}
The dynamical evolution of the separation between the SMBH pairs ($\Delta R_\mathrm{BH}$), along with the orbital parameters ($1/a$ and $e$) for the hard binaries, is shown in Fig.~\ref{fig:triple_evolution}. In models A.1, A.2, A.3, and B the SMBH was simulated along with the progenitor galaxies, while in model C, the black holes in the galaxy already resided in the halo for several Gyr.  

Initially in the A.1 simulation, all three galaxies were separated from one another. During the galactic inspiral phase, at around 300~Myr, galaxy 2 became tidally disrupted and partially merged with galaxy 3. As a result, BH2 and BH3 approached each other very closely and formed a common halo, orbiting together until all three black holes came close together during their first pericenter passage at 465~Myr. The BH2-BH3 binary continued to harden and descended further until around 560~Myr, where it underwent a strong galactic collision with BH1 in a chaotic three-body interaction. This prevented the further formation of the BH2-BH3 binary system. Instead, BH1 descended toward the BH2 under the influence of dynamical friction, and a bound BH1-BH2 binary began to form at 600~Myr. Meanwhile, BH3, the lightest SMBH, was ejected onto a radial orbit around the galactic center at a distance of 100\,-\,800~pc. As the BH1-BH2 binary began to harden, BH3 gradually began to descend toward the inner hard binary, its orbit circularized, eventually resulting in the formation of a stable hierarchical triple system in which BH3 became gravitationally bound to the inner BH1-BH2 binary pair.

In both the A.2 and A.3 simulations, the evolution of the galactic inspiral phase is very similar to that in A.1. However, in simulation A.2, the BH2-BH3 pair continues to harden for an additional 80~Myr compared to A.1, while BH1 approaches more closely toward the galactic center with BH2 until around 625~Myr. At this point, it undergoes a collision as in A.1, forming a bound BH1-BH2 binary. Meanwhile, BH3 is ejected onto a 50\,-\,500~pc orbit around the galactic center. The BH1-BH2 binary hardens slowly compared to A.1. On the other hand, BH3 descends toward the hard binary over the next 250~Myr ultimately forming a hierarchical triple system. This is also the same outcome in A.3, although the three black holes experience a chaotic three-body interaction and remain very close to each other for a short period of 75~Myr until finally separating at 635~Myr. Consequently, the BH1-BH2 pair binds and forms a hard binary, while BH3 settles into a bound outer orbit around the binary, again resulting in a hierarchical triple system.

In simulation B, the merger occurs between two galaxies, where BH1 resides at the center of galaxy 1, while BH2 and BH3 form a close pair in galaxy 2 with orbital separations of 50\,-\,1500~pc. The two galaxies undergo a strong collision and begin to merge after 235~Myr, which separates the close (but unbound) BH2-BH3 pair. BH2 experiences strong dynamical friction since it is closer to BH1 and quickly descends toward the center of the remnant to form a binary pair with BH1. In contrast, BH3 is ejected onto a wide galactic orbit via a slingshot effect at a distance of 1000-3000~pc. The BH1-BH2 pair quickly hardens and forms a bound binary at 320~Myr and eventually merges after 863~Myr. Due to the small orbital period, the direct $N$-body integration slowed down considerably, and the computational effort increased significantly. Hence, we artificially merged the BH1 and BH2 when their separation dropped below $\sim10^{-4}$~kpc and used Eqs.~(\ref{1/a_petermathews}) and (\ref{e_petermathews}) to estimate the merging time. BH3 on the other hand, experiences very little dynamical friction and therefore remains on a wide galactic orbit for the next Gyr.

In simulation C, all the SMBHs reside in the common halo for several Gyr. In this system, BH1, the heaviest black hole, remains in the center. By contrast, BH2 and BH3 -- the lighter black holes, remain in wide galactic orbits with an initial apocenter of 2300~pc around BH1. BH2 approaches the center much more closely, with a pericenter of around 400 pc, while BH3 remains farther away at distances greater than 1300 pc. All the SMBH pairs remain in very close orbits for more than a Gyr, but none gets bound. Slowly after 1100~Myr, due to dynamical friction, BH2 descends towards the center and reaches the influence radius of BH1, thus finally forming a bound BH1-BH2 binary pair after 1300~Myr. By contrast, BH3 does not influence the inner binary much and remains in a stable and wide orbit around the inner pair with almost the same peri- and apocenter. BH1-BH2 hardens soon after this at a constant hardening rate. This hence suggests its eventual merger. 

In all of our simulations with initial triaxial galaxies, the black holes arrange themselves such that the heaviest black hole settles at the center of the merger remnant, while the two lighter black holes wander around it. Eventually, the two heaviest black holes form a bound binary, whereas the fate of the third black hole varies. It either forms a hierarchical triple (A.1, A.2, and A.3) or is never bound and remains on a wide galactic orbit around the inner binary (B and C).

\subsection{Hardening rates and merger times}
\begin{table}[!ht]
\caption{Hardening rates and estimated merger times of the SMBH binaries.}
\label{tab:mergertimes}
\begin{tabular}{>{\centering\arraybackslash} p {2.7 cm }  >{\centering\arraybackslash} p { 0.7 cm} >{\centering\arraybackslash} p {0.7 cm} >{\centering\arraybackslash} p {0.7 cm} >{\centering\arraybackslash} p {0.7 cm}  >{\centering\arraybackslash} p {0.7 cm} }
\toprule
Property& A.1 & A.2 & A.3 & B & C\\
\midrule
 $s_{12}$ & 1.50 & 3.64 & 1.82 & 131.40 & 2.10\\
 $s_{12}^\dag$ & 2.30 & 2.92 & 3.20 & 400.66 & 7.76\\
 $H_{12}$ & 52.30 & 41.52 & 11.3 & 5.75 & 6.68 \\
 $H_{12}^\dag$ & 7.8 & 11.3 & 6.5 & 4.5 & 3.4 \\
 $t_{c,12}\ \text{in Myr}$    & 2863 & 7400 & 3160 & 863* & 13409\\
 $t_{c,12}^\dag\ \text{in Myr}$    & 2680 & 1489 & 1450 & 694* & 2058\\
 \midrule
 $s_{12-3}$  & 0.35 & 0.43 & 0.41 & - & - \\
 $H_{12-3}$ & 4.35 & 1.81 & 0.85 & - & - \\
 $t_{c,12-3}\ \text{in Myr}$ & 67031 & 52893 & 53981 & - & - \\
\bottomrule
\end{tabular}
\tablefoot{*Merger of the binary, which occurred during the simulation. $^\dag$Same parameters for the spherical galaxy models from \cite{2023A&A...678A..11K}. The stellar hardening rates ($s$) are given in kpc$^{-1}$~Myr$^{-1}$, $H$ is dimensionless, and the projected coalescence time $t_c$ is given in Myr following the start of the simulation. The top block shows these quantities for the final inner BH1-BH2 binaries. The bottom block shows these quantities for the final BH3 orbit around the BH1-BH2 remnant, i.e., for the hierarchical triple in A.1, A.2, and A.3.}
\end{table}

After our simulations reached $t_\mathrm{end}$, we estimated the merger time ($t_\mathrm{c,12}$) of the inner BH1-BH2 hard binary using Eqs.~(\ref{1/a_petermathews}) and (\ref{e_petermathews}). We also used this to extrapolate the orbits of the BH1-BH2 hard binary around its center of mass. In all the simulations, the two most massive black holes merge first. The exact merger time differs between each model because every system has a different stellar density and triaxiality parameters, all of which influence the orbital evolution and the stellar hardening rate $H_\mathrm{12}$ of the hard binary. The results are reported in Table~\ref{tab:mergertimes}.

In simulations A.1, A.2, and A.3, where BH3 forms a hierarchical triple with the BH1-BH2 binary, along with the inner binary, we also applied Eqs.~(\ref{1/a_petermathews}) and (\ref{e_petermathews}) to obtain the merging time $t_\mathrm{c,12-3}$ of BH3 around the merged BH1-BH2 remnant. We can clearly see that the inner BH1-BH2 binaries show significantly higher hardening rates than BH3, implying that BH3 takes a significantly longer time to merge, exceeding even Hubble time. 

Models A.1, A.2, and A.3 exhibit similar hardening, consistent with nearly identical dynamical friction phases for the inner BH1-BH2 pair. They coalesce over a few Gyr in line with earlier estimates of merger timescales for SMBHs in massive galaxies \citep{2018A&A...615A..71K,KHB2025}. The inner binary in simulation B merged during the simulation due to stronger dynamical friction and the highest stellar hardening rate (Table~\ref{tab:mergertimes}). In contrast, simulation C shows a smaller hardening parameter compared to the A models due to the low density of its surrounding remnant. As a consequence, the BH1-BH2 binary takes a significantly longer time to reach the $a_{h}$. This results in a merger timescale nearly equivalent to Hubble time.

\subsection{Evolution of the triaxiality} \label{Evolution of the triaxiality}
\begin{figure*}[h]
    \centering
    \includegraphics[width=\textwidth]{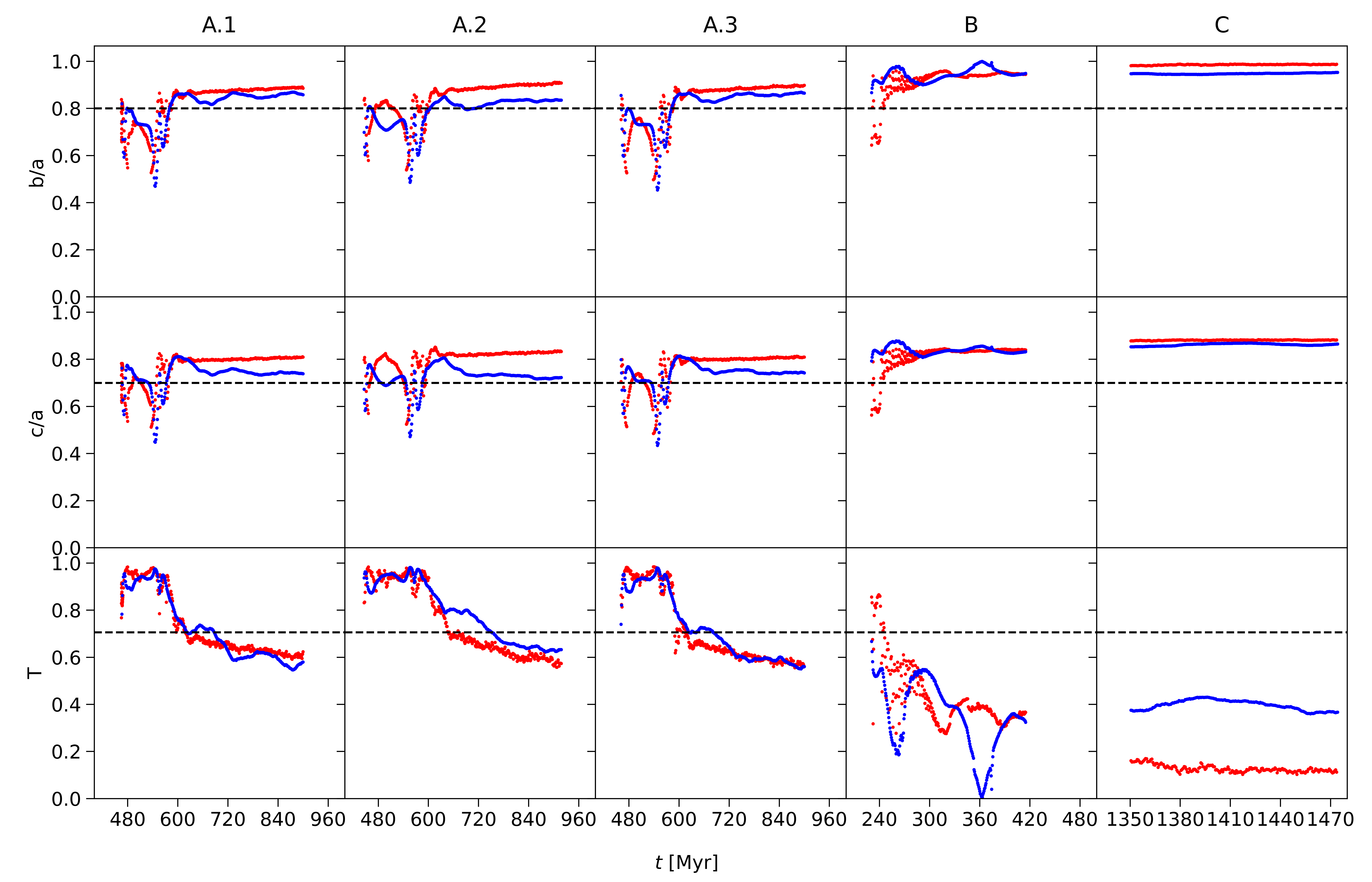}
    \caption{Time evolution of galaxy shape parameters for all the simulations during SMBH evolution with \texttt{$\varphi$-GPU} at a distance of 1~kpc (red) and 5~kpc (blue). Each column corresponds to one simulation, and the rows indicate the axis ratios $\mathrm{b}/\mathrm{a}$ (top), $\mathrm{c}/\mathrm{a}$ (middle), and triaxiality parameter $\mathrm{T}$ (bottom). The dashed black lines correspond to the initial triaxiality values for the stellar component of the initial progenitor galaxies before the simulation. Please note that the time ranges for each model are different.}
    \label{fig:triaxial_evolution}
\end{figure*}
The evolution of the galaxies, together with the merger and sinking of their SMBHs, alters the intrinsic shape of the galactic remnant and can influence the merger timescale of the triple. Using the axis ratios $\mathrm{b}/\mathrm{a}$ and $\mathrm{c}/\mathrm{a}$, as well as the triaxiality parameter $\mathrm{T}$ introduced in Sect.~\ref{Triaxial Galaxies}, we tracked the global structure and time-dependent shape evolution of the stellar remnant in each model during the interaction at three different radii (0.1, 1, and 5 kpc) around the center of mass of the remnant once the galaxies merged (see Figs.~\ref{fig:triaxial_evolution} and \ref{fig:triaxial_evolution_0.1kpc}). We followed the well-known Jacobi eigenvalue algorithm formulation used in \cite{2023A&A...677L...6R} to calculate our triaxial shape parameters. 

All three A-models exhibit qualitatively similar behavior across the three radii. Within the innermost 0.1 kpc (Fig. \ref{fig:triaxial_evolution_0.1kpc}), the axis ratios show dips during the early three-body interaction phase ($\approx$\,450\,-\,650~Myr) when the SMBHs undergo repeated close passages. These events drive the nucleus into strongly triaxial configurations, particularly during phases of enhanced SMBH-SMBH interactions and rapid potential fluctuations \citep{2011ApJ...732...89K, 2011ApJ...732L..26P}. At 1 kpc, the same perturbations continue but are significantly weaker, indicating that the influence of the triaxial structure on the SMBH dynamics decreases rapidly with radius. After the formation of the BH1-BH2 hard binary around 600~Myr, both $\mathrm{b}/\mathrm{a}$ and $\mathrm{c}/\mathrm{a}$ steadily increase, the orbits circularize, and $\mathrm{T}$ slowly declines. This indicates that the remnant relaxes toward a more spherical but slightly oblate structure as the central potential deepens due to binary hardening and stellar scattering. At 5 kpc, only mild variations are observed. The large-scale halo retains its original triaxial configuration and is only weakly affected by the central dynamical evolution. This is in agreement with previous merger simulations, which show that outer halos preserve memory of their initial shapes \citep{1991ApJ...378..496D, 2006MNRAS.367.1781A}. Overall, all the A-models settle into moderately triaxial, slightly oblate remnants with shapes comparable to those of the progenitor galaxies, with differences between A.1, A.2, and A.3 mainly reflecting their varying stellar density profiles.

Simulation B shows the strongest and most rapid structural evolution among all simulations. The violent encounter between the two progenitor galaxies and the preexisting BH2-BH3 pair induces extreme distortions within the central 0.1~kpc resulting in a sharp drop in both the axis ratios during $\approx$ 230\,-\,300~Myr. By contrast, $\mathrm{T}$ increases, indicating a temporarily prolate-to-triaxial nucleus. Once the BH1-BH2 binary forms, the system undergoes a rapid transition toward a more spheroidal configuration. At 1 kpc, the merger produces a clear but less extreme tidal response. The initial prolate signature gradually dissipates, and by $\approx400$~Myr the intermediate region approaches an oblate shape. Even at 5 kpc, the tidal imprint of the merger remains visible, although it fades quickly as the outer halo relaxes with high axis ratios. Hence, the system gradually evolves toward a moderately triaxial shape.

Simulation C shows the weakest shape evolution, consistent with its relatively quiet dynamical history. All SMBHs initially reside within the same halo, and no major galactic merger occurs during the simulated period. As a result, the axis ratios show only gentle, smooth variations at 0.1 kpc. The nucleus remains close to oblate, with the slow inspiral of the BH1-BH2 binary. At 1 kpc, we observe a very high $\mathrm{b}/\mathrm{a}\sim1$. Thus, the shape evolves slightly toward a more spherical, weakly oblate disky configuration as the black holes sink toward the center, but without the strong tidal disruptions seen in the previous models. At 5 kpc, the outer halo remains essentially unchanged and preserves the initial triaxial geometry.

Across all simulations, the degree of triaxiality and the magnitude of structural evolution strongly depend on radius. The innermost 0.1-1 kpc, where the SMBHs interact and form binaries, is most sensitive to dynamical perturbations and shows the largest variations in shape. In contrast, the outer 5~kpc halo is largely unaffected and retains memory of the initial triaxial structure. A.1, A.2, and A.3 experience repeated but moderate distortions, while B undergoes a single violent transformation followed by rapid relaxation, and C remains dynamically quiet with only mild evolution and finally circularizes. Hence, although the triaxial structure evolves during the initial interaction phase, once the binary settles, the remnant in all simulations retains a shape that does not exhibit any significant departure from the original triaxial structure of the progenitor galaxies.

\subsection{Eccentricity evolution} \label{Eccentricity evolution}
The eccentricity evolution exhibits qualitatively similar trends across the simulations. In models A.1 and A.3, the inner BH1-BH2 binary reaches very high eccentricities ($e \gtrsim 0.9$), whereas in A.2 the slower infall of the black holes toward the center results in a lower eccentricity of $e \sim 0.5$ for the inner binary. Previous studies have shown that the eccentricity of the inner orbit seems to constantly decrease during the dynamical friction phase until the formation of the hard binary \citep{2022MNRAS.511.4753G}. This behavior is particularly seen in simulation B. Here, the SMBHs experience a strong dynamical friction from the stellar particles, leading to high stellar hardening. This allows the inner binary to quickly harden, decreasing the eccentricity of the binary as it moves toward coalescence and finally merges through circularization. On the other hand, the SMBH binary in simulation C experiences a relatively longer dynamical friction phase and has a low eccentricity as compared to all the models in system A, ultimately resulting in the longest merger timescale. 

Despite the very high eccentricities reached in models A.1 and A.3, the late-time evolution of the hierarchical triple shows only modest oscillations in the eccentricity of the inner binary. We do not attribute these variations to the Kozai-Lidov-von Zeipel (KLZ) effect \citep{1962AJ.....67..591K, 1962P&SS....9..719L, 2016ARA&A..54..441N}, since this mechanism assumes third perturber is a test particle, as in EMRIs \citep{2022MNRAS.516.1959M}. In our simulation, however, this is not the case, because the third black hole has a comparable mass. This leads to dynamics dominated by direct three-body interactions, characterized by strong, chaotic gravitational exchanges \citep{2007MNRAS.377..957H, 2018MNRAS.477.3910B} rather than a hierarchical framework. Recent work by \cite{2024MNRAS.52710705H} has also verified that KLZ oscillations are short-lived and rare in triple SMBHs. 

\section{Discussion and conclusion} \label{Discussion and conclusions}
\begin{figure*}[h]
   \centering
    \includegraphics[width=0.9\textwidth]{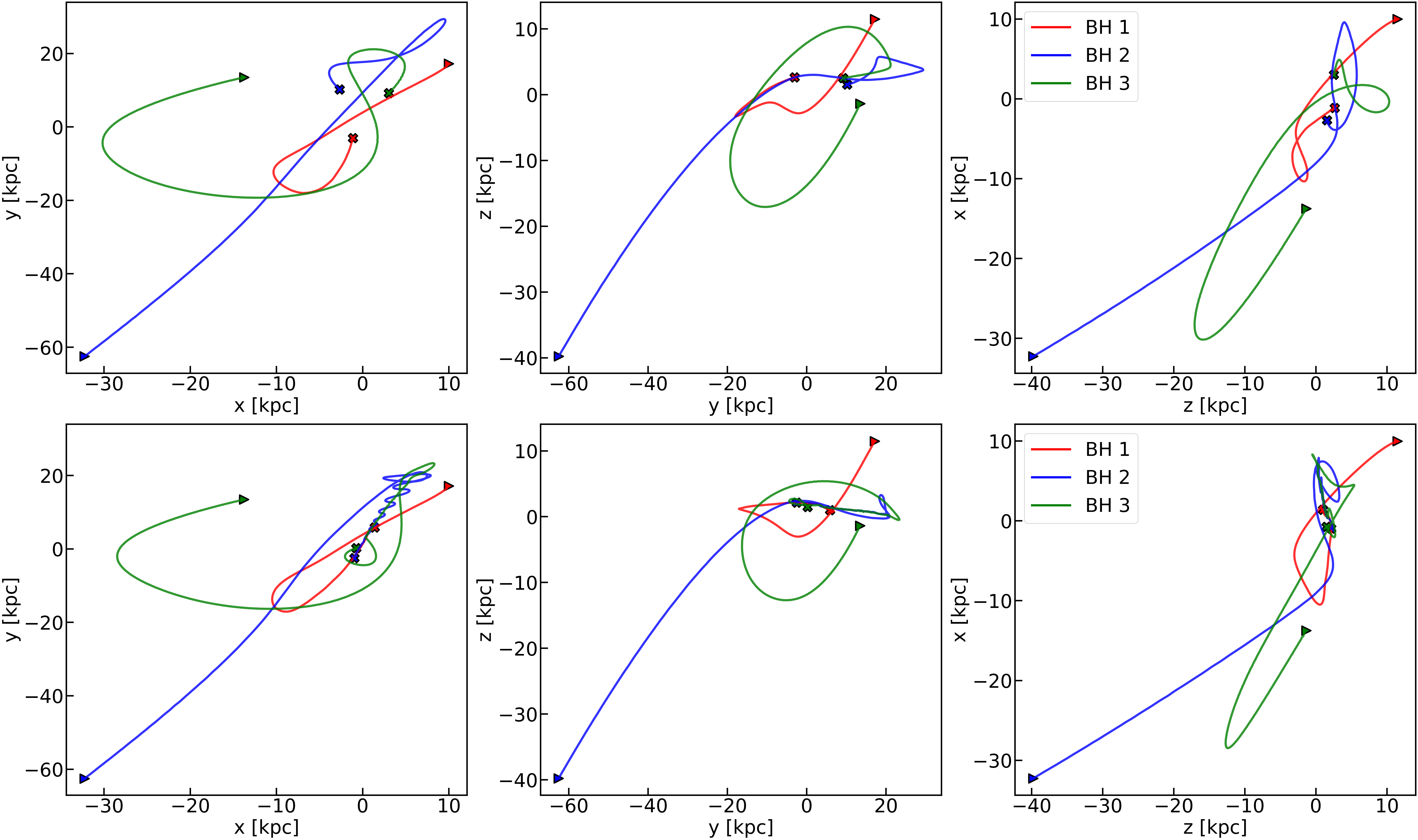}
   \caption{Orbital evolution of the three SMBHs in model A.3, simulated using \texttt{Bonsai2}, showing the continuous exchange of the orbits due to three-body encounters in spherical galaxies (top) and triaxial galaxies (bottom) between $t=0$ to $t=500$~Myr. The starting and ending points are marked with a triangle and a cross, respectively. Please note that the scales of the $x$-, $y$-, and $z$-axis are different.}
    \label{fig:smbh_orbits}
\end{figure*}

In this paper, we followed the dynamical evolution of SMBH triple systems extracted from the ROMULUS25 cosmological simulation and resimulated them in self-consistent triaxial stellar potentials using gravitodynamical $N$-body simulations. Unlike earlier studies based on spherical galaxy models by \cite{2021A&A...649A..41A} and \cite{2023A&A...678A..11K}, our initial conditions preserve the intrinsic non-axisymmetric structure of the host galaxies and hence capture more realistic orbital structure and dynamics of the merger remnants. This allowed us to quantify the joint impact of triaxiality and triple SMBH interactions on hardening rates, eccentricity evolution, and the merger timescales.

In \cite{2023A&A...678A..11K}, the ROMULUS25 galaxies were modeled as spherical systems, an assumption that directly influences the inspiral phase and the subsequent orbital evolution of the SMBHs. In such idealized configurations, the gravitational potential is spherical, limiting the range of stars which are on centrophilic stellar orbits that can efficiently extract energy and angular momentum from the SMBH system \citep{2001ApJ...563...34M, 2002MNRAS.331..935Y}. However, it is well established that galaxy mergers naturally destroy initial sphericity. The dynamical interactions between the merging galaxies and their SMBHs redistribute stellar orbits anisotropically, inducing triaxiality and flattening in the gravitational potential \citep{2011ApJ...732...89K, 2018MNRAS.477.2310B}. Our results are also consistent with \cite{2018MNRAS.477.2310B} in terms of the merging timescales of the binary and the triaxial shape parameters of the non-axisymmetric merger remnants.

The original galaxies in the ROMULUS25 simulation on the other hand are not spherical but instead exhibit intrinsic triaxiality. Such nonspherical potentials sustain a larger population of centrophilic orbits and increase the system's susceptibility to tidal forces, hence influencing stellar hardening and facilitating continued SMBH orbital decay. As a result, the orbital evolution of the galactic remnant and, consequently, the SMBH dynamics can differ significantly from the idealized spherical case \citep{2006ApJ...642L..21B, 2013ApJ...773..100K, 2017MNRAS.464.2301G}. We verified this by comparing the orbital evolution of the triples present in initially spherical potentials \citep[as done in ][]{2023A&A...678A..11K} and a triaxial potential with the same initial conditions, as shown in Fig.~\ref{fig:smbh_orbits}. 

In the spherical case, the black holes followed relatively smooth and regular trajectories during the early stages of evolution. In the initial dynamical friction phase, the motion of the SMBHs is predominantly confined to well-defined, rosette-like orbits (see Fig.~\ref{fig:smbh_orbits}) supported by the isotropic mass distribution \citep{Binney2008}. However, as the system evolves and the galaxies undergo dynamical mixing alongside multiple SMBHs, the global symmetry is progressively perturbed. Hence, strong gravitational interactions and energy exchange induce departures from perfect sphericity, thereby modifying the orbital structure. In contrast, the triaxial model exhibited intrinsically more complex orbital behavior from the start. The absence of rotational symmetry results in more irregular and spatially extended centrophilic orbits with stronger three-body interactions toward the central region. As the black holes come very close, they experience repeated close pericenter passages, which change the orbits and hardening of the triple system.

Furthermore, our results show that triaxiality and SMBH multiplicity act in a complementary manner. While triaxiality alone is known to prevent long-term loss cone depletion by allowing a steady flux of stars on centrophilic orbits and solving the final parsec problem \citep{2004ApJ...606..788M, 2006ApJ...642L..21B, 2011ApJ...732...89K, 2013ApJ...773..100K, 2017MNRAS.464.3131K, 2017MNRAS.464.2301G}, the presence of a third SMBH introduces repeated phases of strong three-body interactions, orbital exchanges, and transient hierarchical configurations, as seen in model A. 

In Table~\ref{tab:mergertimes}, we compared our hardening rates with those obtained by \cite{2023A&A...678A..11K}. It has already been shown that binaries present in a spherical potential experience reduced hardening compared to triaxial configurations. We find that our hardening parameter $H_{12}$ was relatively higher than that obtained in the spherical galaxies, but largely consistent within a factor of $\sim2$, with the theoretical values of $15\lesssim H\lesssim 20$ obtained by three-body scattering experiments \citep{2006ApJ...651..392S, 2020MNRAS.493L.114B}. This was also in agreement with realistic SMBH binary systems simulated using direct $N$-body \citep{2015MNRAS.454L..66S} and Monte Carlo \citep{2019MNRAS.484.2851L} recipes. Nevertheless, despite a higher $H_{12}$ in all triaxial configurations, the estimated merger timescale of the inner binary is usually longer than in the spherical cases.

During dynamical evolution, we observe that the heavier BH1-BH2 binary merges first, while the lightest black hole, if at all, usually takes much longer to merge. This is similar to earlier studies on triple systems \citep{2006ApJ...651.1059I, 2007MNRAS.377..957H, 2021ApJ...912L..20M} and fully consistent with the recent result of \cite{2023A&A...678A..11K}. Although the hard inner binary coalesces within Hubble time in all models, the merger timescales differ substantially.

In models A.1, A.2, and A.3, a stable hierarchical triple system forms, with the heavier BH1-BH2 binary merging first, after the three SMBHs directly undergo a chaotic three-body interaction with each other, and the BH3 orbits around this binary. In A.1 and A.3, the binary exchange between the SMBHs takes place earlier, thus allowing BH1-BH2 to settle down into a highly eccentric orbit after the separation; hence, the binary merges within a fraction of the Hubble time. On the other hand in A.2, due to a delay in  dynamical friction, the binary takes a bit longer to harden and settles in a less eccentric orbit, thus taking a longer time to merge than A.1 and A.3. The most rapid coalescence is observed in model B, where BH2 is initially located much closer to BH1; it quickly hardens from strong dynamical friction and merges in less than a Gyr. In contrast, in model C, where all the galaxies have already merged, all three SMBHs undergo repeated close three-body interactions for almost a Gyr. This weakens the central density cusp of the system, slowing dynamical friction. Hence, the inner BH1-BH2 binary takes more than a Gyr to reach the influence radius, thus delaying the merger, to a timescale approaching the Hubble time.
\begin{figure*}[h]
    \centering
    \includegraphics[width=\textwidth]{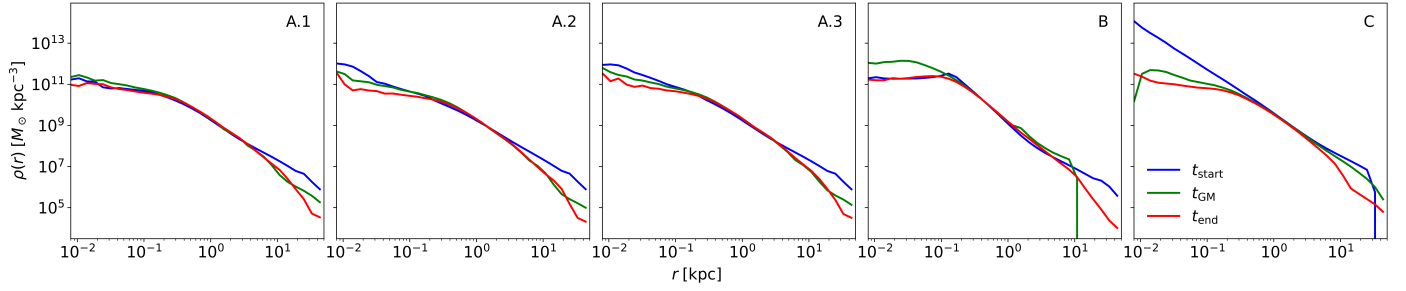}
    \caption{ Volume density profiles of the five models at the beginning ($t_{\rm start}$, in blue), after the first pair of galaxies merges at $t_{\rm GM}$\,=\,590, 625, 630, 230, 1100~Myr for models A.1, A.2, A.3, B, and C, respectively (in green), and at the end ($t_{\rm end}$, in red) of the simulation. }
    \label{fig:density_profile}
\end{figure*}

These variations in the merger timescales are closely linked to the structural evolution of the central density profile. We attribute the longer merging time to the reduced central stellar density of the merger remnant, as shown in Fig. \ref{fig:density_profile}, where the density profiles are compared both at the beginning -- after the first pair of galaxies merges ($t_{\rm GM}$) -- and at the end of the simulation. In the triaxial system, strong internal mixing and mass redistribution of the stellar particles during the galaxy interaction lead to a dynamically more diffuse central region. This is similar to that likewise found in the spherical galaxy, where the bound triple SMBH binaries remove matter inside the sphere of influence, thus flattening the density profile. This "binary scouring" is not only a result of SMBH scattering \citep{2001ApJ...563...34M, 2007MNRAS.377..957H, 2014ApJ...782...39T, 2024ApJ...974..204K} but is also fundamentally linked to the self-consistency of the triaxial potential. As found in studies of elliptical galaxy models, high central density regions with a central SMBH or a steep stellar density cusp disrupt the stable box orbits necessary to maintain triaxiality, often forcing the center to become more axisymmetric or spherical \citep{1996ApJ...460..136M, 1997ApJ...486..102M}. 

In our triaxial models with a central density cusp ($\rho \propto r^{-\gamma}$), a significant fraction of orbits become chaotic due to close three-body interactions. As demonstrated in an early work by \cite{1985MNRAS.216..467G}, these stochastic orbits undergo slow diffusion in phase space, effectively smoothing out the central density and acting against maintaining a steep cusp \citep{2001sdcm.conf..420S, 2001ApJ...549..192P}. Thus, weakening the central cusp ($\gamma\lesssim1$) for our models either due to scouring (B and C) or chaotic SMBH three-body interactions (A.1-A.3) consequently decreases the density in inner galactic regions \citep{1987IAUS..127..241G, 1996ApJ...471...82M}. This lower central density naturally results in a smaller physical hardening rate $s_{12}$ \citep{1996NewA....1...35Q}, thereby prolonging the merger timescale of the triaxial galaxies. This effect is particularly evident in models B and C, where the reduced central stellar density results in a $s_\mathrm{12}$ that is a factor of $\sim$3 smaller than in the corresponding spherical models (see Table~\ref{tab:mergertimes}). Lastly, the longer coalescence times are also linked to the smaller mass ratios of BH3 to the BH1-BH2 remnants, which prolongs the GW emission phase.

Cuspy density profiles can enhance orbital decay and accelerate binary hardening through efficient dynamical friction \citep{2008ApJ...678..780G, 2011MNRAS.411..653J, 2012ApJ...745...83A, 2020MNRAS.493.3676O}, hence increasing the hardening rate of the SMBH binary \citep{2016ApJ...828...73K, 2018A&A...615A..71K}. Following this argument, \cite{2023A&A...678A..11K} claimed that the hardening efficiency for the outer orbits in the cuspy triple system could increase following the merger of the inner binary. We find no evidence of this in triaxial models B and C. In our case, the orbits of BH3 experience almost no dynamical friction due to the remnant's low central density, as a result of chaotic three-body interactions, and thus remain on a wide orbit.

Finally, in triple systems such as ours, the interaction of the heavier binary with the lighter SMBH produces characteristic fluctuations in the instantaneous hardening rate, particularly during close three-body encounters. These episodes temporarily increase the rate of energy extraction and the exchange of angular momentum from the inner binary and can trigger rapid eccentricity growth, as seen in models A.1, A.3, and B. This may accelerate the  merger of the binary but usually lacks a KLZ-like oscillation due to relativistic periastron precession.

We find that the global triaxial structure of the host galaxy is largely preserved outside the central kiloparsec, remaining stable within a 20\% margin across all our models. Conversely, the innermost regions ($\sim$0.1\,-\,1 kpc) within the sphere of influence undergo the most rapid shape evolution during the chaotic three-body encounters of the SMBHs (see Appendix \ref{Triaxiality at smaller radii}). Consequently, at large radii, the models maintain their triaxiality \citep{1985MNRAS.216..467G, 2002ApJ...567..817H, 2004ApJ...606..774P}. This radial dependence is a key outcome of our simulations. It implies that the regions most relevant to the formation of a hard SMBH binary are also those most susceptible to non-axisymmetric perturbations. The hierarchical models mostly circularize toward a moderately triaxial structure. In contrast, the remnants in models B and C, where BH3 remains in a wide orbit far from the influence radius, circularize toward a nearly spherical structure. Thus, the SMBHs evolve within a time-dependent potential where the local triaxial inner structure is actively reshaped by the dynamics of the triple system.

Our results show that in all the triaxial models, the two heaviest SMBHs form a hardening binary which merges first. The fate of the lightest black hole, however, is largely dependent on our initial conditions, in which it either forms a hierarchical triple or remains on a wide galactic orbit without close interaction with the hard binary, similar to the spherical cases \citep{2023A&A...678A..11K}. Hence, we conclude that the initial triaxiality of the galaxies does not have a very strong effect on the final outcome and the merging timescale of the inner binary in the triple systems. The hardening of the outer binary is usually delayed beyond the Hubble time due to binary scouring by the inner hard binary, which reduces dynamical friction. We ignored any hydrodynamical effects in our simulations, in which the infalling gas might influence the central density, dynamical friction, and the hardening rates of the SMBHs. Nevertheless, various zoom-in cosmological simulations of galaxy mergers with multiple SMBH binaries have already suggested that neglecting hydrodynamics does not change the final outcome of the triple drastically \citep{2019MNRAS.486.4044B, 2021ApJ...912L..20M, 2022ApJ...929..167M, 2022MNRAS.510..531C}.

We conclude that high-resolution, gravitodynamical $N$-body simulations in more realistic triaxial galaxies are essential for capturing the complex dynamics of triple SMBHs, which are often unresolved in wider scale cosmological simulations. Our results demonstrate that galactic triaxiality and complex three-body interactions are primary drivers of orbital evolution and coalescence in triple SMBH systems. These factors govern the efficiency of dynamical friction and the subsequent hardening rates of the binaries, extending beyond the scope of standard binary models. New statistics from various cosmological simulations reveal that a large number of binaries found in these simulations undergo mergers to form triple SMBH systems \citep{2024MNRAS.527.7424S, 2025ApJ...993..222S}. Hence, future $N$-body simulations of these SMBH triples should be performed using more realistic galaxy shapes rather than using spherical symmetry. Using a hybrid technique, for example, via particle splitting from cosmological simulations may prove to be vital for the detailed dynamical evolution of multi-SMBH systems \citep{2002MNRAS.330..129K, 2007MNRAS.379..956D, 2007Sci...316.1874M, 2010Natur.466.1082M, 2015MNRAS.449..494R, 2016ApJ...828...73K}. 

Several questions still remain regarding how these stable hierarchical triples interact in the presence of nuclear clusters or central gaseous disks, and how they affect long-term hardening and merger timescale of the inner binary, as well as the fate of the lighter black hole. Future work must also prioritize the inclusion of a wider triaxiality range along with three-body PN terms, velocity recoils, and orbital spin evolution \citep{2018MNRAS.477.3910B, 2021ApJ...912L..20M, 2022ApJ...929..167M}. Such detailed modeling is essential for moving beyond the secular approximation and for fully understanding the behavior of multiple SMBH systems as they occur within the hierarchical framework of galaxy formation.

\begin{acknowledgements}

We thank Michael Tremmel for providing the ROMULUS25 simulation data, and Kai Wu, and Marcelo C. Vergara for their helpful discussions. We thank the anonymous referee for constructive feedback that helped improve the clarity and quality of this manuscript. The authors gratefully acknowledge the Gauss Center for Supercomputing eV\footnote{www.gauss-centre.eu} for funding this project by providing computing time through the John von Neumann Institute for Computing (NIC) on the GCS Supercomputers JUWELS Booster and JUPITER Booster at Jülich Supercomputing Center (JSC)\footnote{www.fz-juelich.de/en/jsc}. The authors acknowledge support by the High Performance and Cloud Computing Group at the Zentrum für Datenverarbeitung of the University of Tübingen, the state of Baden-Württemberg through bwHPC through grant no. INST 37/1159-1 FUGG, and the data storage service SDS@hd supported by the Ministry of Science, Research and the Arts Baden-Württemberg (MWK) and the German Research Foundation (DFG) through grant no. INST 35/1503-1 FUGG. NS acknowledges the support of a doctoral scholarship from the German Academic Exchange Service (DAAD) through grant No. 91864951. PB and MS are grateful for the support from the special program of the Polish Academy of Sciences and the U.S. National Academy of Sciences under the Long-term program to support Ukrainian research teams grant No.~PAN.BFB.S.BWZ.329.022.2023. We also gratefully acknowledge the Polish high-performance computing infrastructure PLGrid (HPC Center: ACK Cyfronet AGH) for providing computer facilities and support within computational grant No.~ PLG/2026/019243. This material is based upon work supported by Tamkeen under the NYU Abu Dhabi Research Institute grant CASS.

\end{acknowledgements}

%
%
\bibliographystyle{aa}

\bibliography{references}  

\begin{appendix}
\onecolumn
\section{Triaxiality at smaller radii} \label{Triaxiality at smaller radii}
\begin{figure}[h!]
    \centering
    \includegraphics[width=\textwidth]{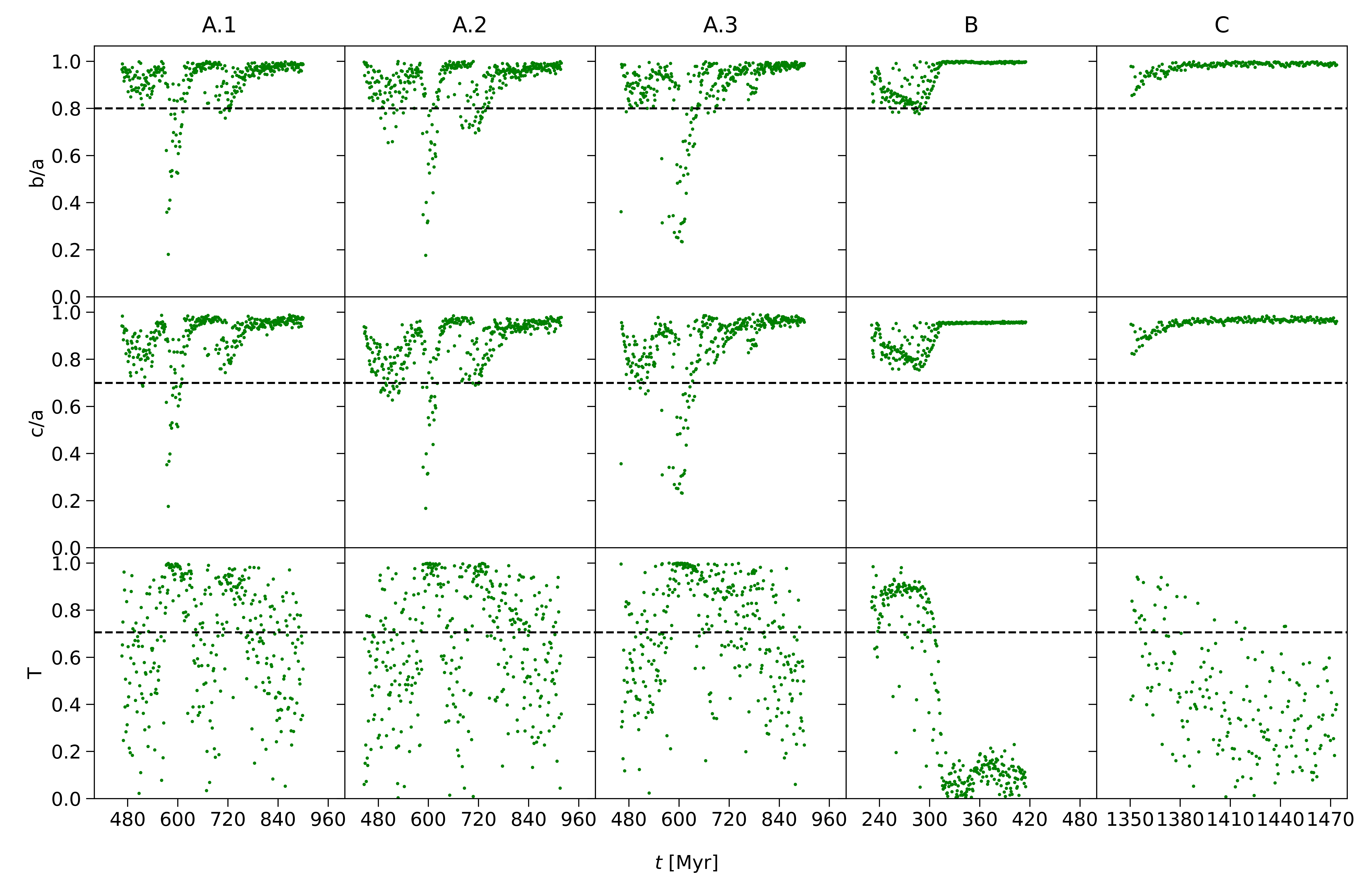}
    \caption{\centering Same as Fig.~\ref{fig:triaxial_evolution} but at a distance of 0.1 kpc from the remnant's center of mass.}
    \label{fig:triaxial_evolution_0.1kpc}
\end{figure}

To assess the robustness of our shape measurements, we compute the triaxiality parameter at a smaller galactocentric distance of 0.1\,kpc close to the influence radius. The time evolution of the triaxiality at this radius is shown in Fig.~\ref{fig:triaxial_evolution_0.1kpc}.

Compared to the larger radii discussed in the section \ref{Evolution of the triaxiality}, the inner region exhibits significantly stronger fluctuations. This increased scatter is expected, as 0.1\,kpc lies close to the gravitational influence radius of the SMBH binary. In this regime, the stellar distribution is strongly affected by both the preceding galaxy merger and the subsequent dynamical interaction with the hardening binary. Stellar scattering, mass redistribution, and the depletion of centrophilic orbits lead to rapid structural changes in the inner central region.

Although the instantaneous values of the triaxiality parameter show enhanced variability at this radius, the overall qualitative trends remain consistent with those observed at larger scales. In particular, the inner 0.1\,kpc region exhibits the strongest dynamical response to the SMBH interactions, while the relative ordering between the models is still preserved.  

\end{appendix}

\end{document}